**TITLE:**

An image-guided high-precision research platform for ultra-high dose rate spinal cord toxicity studies

**AUTHORS:**

Banghao Zhou[1], Lixiang Guo[1], Yi-Chun Tsai[1], Albert van der Kogel[2], John Wong[3], Iulian Iordachita[4], Kai Jiang[5], Weiguo Lu[5], Paul Medin[5], and Ken Kang-Hsin Wang[1,5] *

[1] Biomedical Imaging and Radiation Technology Laboratory (BIRTLab), Department of Radiation Oncology, University of Texas Southwestern Medical Center, Dallas, TX, United States of America

[2] Department of Human Oncology, School of Medicine and Public Health, University of Wisconsin, Madison, WI, United States of America

[3] Department of Radiation Oncology and Molecular Radiation Sciences, School of Medicine, Johns Hopkins University, Baltimore, MD, United States of America

[4] Department of Mechanical Engineering, Whiting School of Engineering, Johns Hopkins University, Baltimore, MD, United States of America

[5] Department of Radiation Oncology, University of Texas Southwestern Medical Center, Dallas, TX, United States of America

**\*CORRESPONDENCE:**

Ken Kang-Hsin Wang, PhD

Department of Radiation Oncology, University of Texas Southwestern Medical Center, 2201 Inwood Road, Dallas, TX, 75235, USA.

Tel: 614-282-0859

Email: kang-hsin.wang@utsouthwestern.edu

**ABSTRACT**

**Objective:** While FLASH radiotherapy (FLASH-RT) is recognized for short-term normal tissue sparing, its durability in late-responding organs remains uncertain, limiting clinical adoption. With its clinical importance and steep dose-response, the spinal cord is an ideal model for evaluating FLASH effect on late toxicity. This work introduces a robust image-guided research platform for high-precision irradiation at both conventional (CONV) and ultra-high dose rates (UHDR) to enable FLASH late toxicity studies using a rat spinal cord model.

**Approach:** A modified LINAC was employed to irradiate the C1–T2 rat spinal cord with 18 MeV UHDR and CONV beams. A custom rat immobilization device, a portable X-ray imaging system, and an ion-chamber-based UHDR output monitoring system were integrated to ensure accurate C1–T2 localization and precise dose delivery. A Monte Carlo (MC) dose engine was developed to provide accurate dosimetry and support the interpretation of in vivo results. Scintillator measurements at UHDR were performed within the spinal cord to verify MC results and the precision of our platform.

**Results**: We achieved submillimeter C1–T2 setup accuracy and maintained submillimeter intrafraction motion. Ion chamber readings showed linear correlation with UHDR output ($R^2 = 1$). MC calculations indicated uniform irradiation (<5% non-uniformity) along the central ~13 mm cord, avoiding dose-volume effect. Our CONV beam exhibited distribution close to that of the UHDR beam, with difference < 3%, isolating dose rate as the only variable. Scintillator-measured dose agreed with MC within 4%, with a 100% gamma passing rate (2%/2 mm), confirming both MC accuracy and the platform's high-precision delivery.

**Significance**: We developed the first comprehensive, image-guided preclinical platform for accurate UHDR and CONV irradiation to investigate FLASH-mitigated spinal cord toxicity in rats. This work thus establishes a robust foundation for systematic evaluation of the FLASH effect in late-responding organs and for determining clinical applicability of FLASH-RT.

## 1. INTRODUCTION

FLASH radiotherapy (FLASH-RT), characterized by ultra-high dose rates (UHDR, >40 Gy/s), offers transformative potential to reduce normal tissue toxicity without compromising tumor control, known as the FLASH effect[1]. Despite promising preclinical evidence, key barriers hinder clinical translation of FLASH-RT. Most studies focus on acute toxicity, while its effects on sparing late-responding tissues remain largely unexamined[2–4].

The spinal cord, frequently constituting the most critical late-responding organ, has a well described steep dose-response curve. Radiation exceeding its tolerance significantly increases the risk of myelopathy, leading to paresis, which can be easily discerned[5]. Its dose-limiting nature and clinical relevance make the spinal cord an ideal model for investigating whether FLASH effect sustain in late-responding organs[6].

Numerous studies have characterized spinal cord responses to conventional dose rate radiotherapy (CONV-RT) across various animal models, with the vast majority performed in rats. Studies using rats have generated a large body of clinically relevant data on the influence of dose, fractionation, re-irradiation, and volume on the tolerance of the spinal cord[5]. However, comparable data under FLASH-RT conditions remain unavailable, and no dedicated platform has been established to enable precise spinal cord irradiation in rats at both UHDR and CONV dose rates.

To address this critical unmet need, we propose to use rats as a clinically relevant model to investigate whether FLASH-RT can mitigate radiation-induced spinal cord injury. To ensure the robust and reproducibility of the proposed study, it is essential to employ a high precision research platform capable of accurately delivering both UHDR and CONV irradiation, under comparable beam characteristics to the rat spinal cord, while also enabling precise in vivo dosimetry. Previous works by Bijl et al.[7] showed that when the irradiated spinal cord length was < 8 mm, the median effective dose ($ED_{50}$, the dose causing paresis in 50% of animals) increased dramatically with decreasing length, known as the dose-volume effect. Consequently, many groups have adopted the cervico-thoracic (C1–T2) region (>20 mm in length) for spinal cord toxicity studies to avoid this volume effect. To this end, we have developed a LINAC-based platform that utilizes 18 MeV electrons to uniformly irradiate the C1–T2 region of the rat spinal cord at both UHDR and CONV modes.

Given the rapid delivery of the UHDR beam (~2 Gy per pulse) and steep dose-response curve of the spinal cord[7], achieving precise dose delivery and dosimetry is essential to determine reproducible dose-response relationships. Specifically, an external pulse control system was implemented to control LINAC pulse delivery at UHDR. Moreover, an ionization chamber (IC) positioned beneath the electron applicator was utilized to monitor Bremsstrahlung radiation in real-time, serving as a surrogate for UHDR output monitoring. To achieve precise animal setup and spinal cord localization for irradiation, we designed a custom rat immobilization device integrated with a portable X-ray imaging system, allowing accurate and reproducible localization of the spine under image guidance.

We aim to evaluate the superior normal tissue sparing of FLASH-RT by conducting studies in which dose rate is the only variable. Achieving this requires similar beam characteristics between the UHDR and CONV beams, as well as an accurate dose engine to generate dosimetric plans for evaluating both 18 MeV UHDR and CONV spinal cord irradiation. We developed a Geant4-GAMOS Monte Carlo (MC) dose engine that precisely models the electron beam. Our dose engine was also employed to quantify electron dose inhomogeneities and guide the experiment design to avoid the dose-volume effects, which could confound the assessment of the FLASH effect in sparing spinal cord toxicity. We further validated the MC-calculated FLASH dose and dose rates using image-guided scintillator measurements sampled at 1000 Hz within the spinal cord.

This work presents the first image-guided FLASH platform enabling precise rat spinal cord irradiation at both UHDR and CONV dose rates, providing a valuable tool to evaluate FLASH-RT's potential to reduce late toxicity and support safe clinical translation.

## 2. METHODS AND MATERIALS

### 2.1 Irradiation setup

A Varian 21EX LINAC (Varian Medical Systems, Palo Alto, CA) was modified to deliver 18 MeV electron beams for rat spinal cord irradiation at both UHDR and CONV dose rates. A posterior-to-anterior (P–A) shoot-through geometry, a 2×1 cm² field size, and a 100 cm source-to-surface distance (SSD) were employed to enable uniform coverage of the C1–T2 spinal cord segment. UHDR delivery was controlled using an external pulse control system, enabling modulation of pulses number[8]. For CONV mode, standard monitor unit (MU) control was used. Female Sprague Dawley rats (aged 9–10 weeks, Charles River Laboratories, Wilmington, MA) under isoflurane anesthesia were used for the irradiation studies in accordance with the

Institutional Animal Care and Use Committee (IACUC) at The University of Texas Southwestern Medical Center.

## 2.2 Animal positioning and target localization

To ensure reproducible animal positioning and accurate target localization, a custom 3D-printed rat immobilization device (Fig. 1a), along with a portable X-ray system (DX3000, 65 kV, 2 mA, Dexcowin, Seoul, Republic of Korea), was developed. The immobilization device incorporated stereotaxic ear bars and lateral body supports to stabilize the head, neck, and torso during irradiation. It also featured three parallel embedded radiopaque rulers to facilitate vertebral localization: one central ruler positioned beneath the animal for imaging, and two lateral rulers used for aligning the device with the LINAC light field crosshairs.

To perform target localization, PA X-ray imaging was performed using the X-ray unit with an array detector (Woodpecker Medical Instrument, Guilin, China) placed beneath the immobilization device (Fig. 1b). Along the left–right (L–R) direction, the C1–T2 spine was aligned with the distal ends of the horizontal marker lines on the central ruler, corresponding to the midline of the immobilization device (yellow dashed line in Fig. 1c). In the superior–inferior (S–I) direction, the spine was fully straightened, and the T2 vertebra was first identified on the central ruler (e.g., at 30 mm in Fig. 1c) as it can be easily discerned. The target point, defined as the transverse center of the spinal cord located 10 mm cranial to T2, was then identified on the same ruler (e.g., at 20 mm in Fig. 1c). This position was also marked on the two lateral rulers, external referencing of the target point for alignment. This target point lies near the midpoint of the ~20 mm long C1–T2 spinal segment and was designed to be aligned with the beam's central axis (CAX) for irradiation, ensuring full coverage of the C1–T2 segment within the 2×1 cm² field. Final alignment of the spinal cord target point with the radiation center was achieved by aligning the immobilization device to the LINAC light field crosshairs (Fig. 1d), using the lateral rulers and midline marker as guides.

To assess the accuracy and reproducibility of the setup, 8 control, non-irradiated rats underwent the 2D X-ray guided setup process described above, followed by cone beam computed tomography (CBCT) imaging in a nearby small animal irradiator (X-RAD 225, Precision X-Ray, Madison, CT). The 2D imaging-based localization was validated against CBCT imaging. Because the spine was fully straightened and aligned with the midline of the immobilization device, and the target point could be readily determined relative to the T2 vertebra, the T2 position was used as a representative landmark to evaluate C1–T2 alignment. The T2 position identified on the CBCT image was compared with that determined on the 2D X-ray image to quantify the accuracy of our setup process for C1–T2 localization.

To further assess potential intra-fraction motion of the spinal cord, 12 irradiated rats underwent an additional post-irradiation X-ray imaging and for each rat, the T2 vertebra locations identified on the pre- and post-irradiation images were compared to evaluate any positional changes.

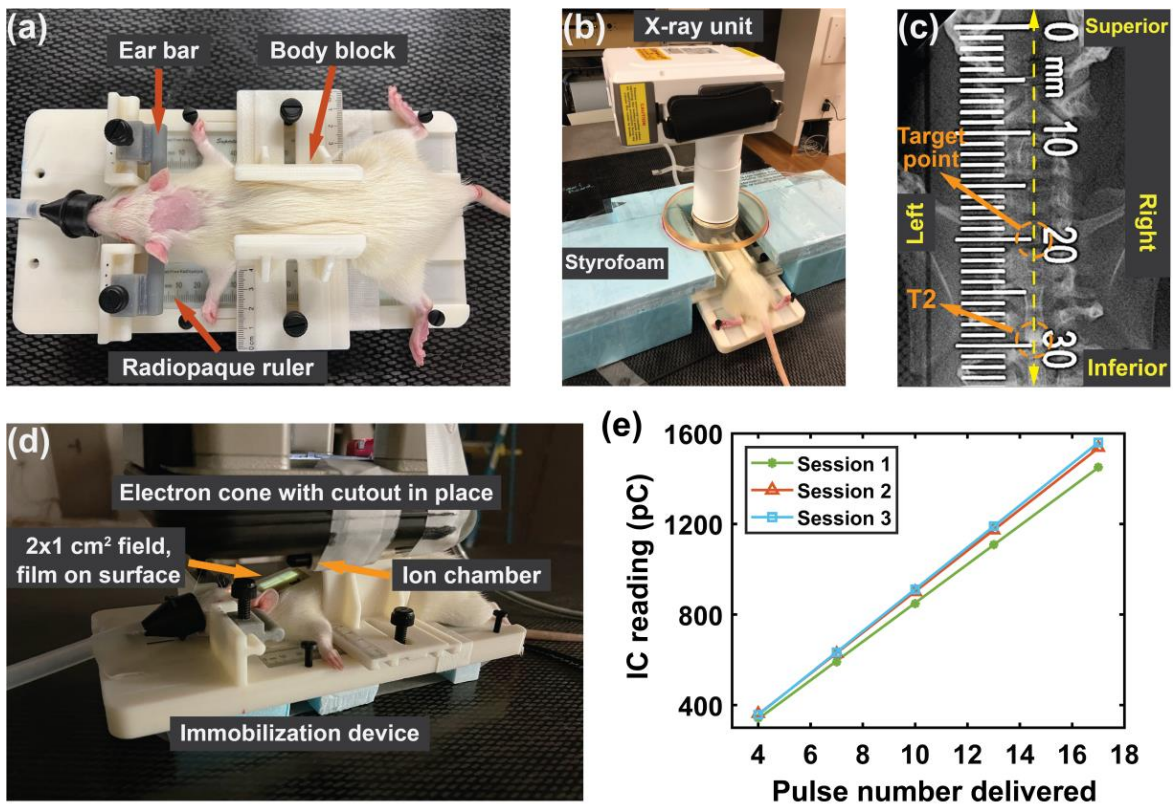

**Figure 1.** (a) Custom rat immobilization device. (b) Setup for X-ray imaging. (c) A representative X-ray image showing identification of the T2 vertebra and corresponding target point. (d) A rat positioned in the immobilization device for UHDR irradiation. The electron cone features a 2×1 cm² cutout for C1–T2 irradiation. For each FLASH group, 3 rats were randomly selected to perform a sanity check of the surface dose rate using the film. A CC13 ion chamber was attached to the cone to measure the Bremsstrahlung radiation, enabling real-time UHDR output monitoring. **(e)** Ion chamber readings linearly correlated with the number of delivered pulses ($R^2 = 1$) across 3 independent measurement sessions conducted over 2 weeks. The within-session standard deviation is smaller than symbol that used in the figure.

## 2.3 UHDR output monitoring

As the LINAC's dose servo system was disabled to enable UHDR delivery[8,9], the inherent beam output regulation was no longer available[10]. To ensure accurate dose delivery, UHDR output was externally monitored to account for any output fluctuations during irradiation for each rat.

This was accomplished by a two-step approach. First, daily absolute dose per pulse (DPP) was determined using Gafchromic EBT-XD films (Ashland, Bridgewater, NJ) irradiated under a reference condition (1 cm depth in solid water, 2×1 cm² field size, 100 cm SSD). Next, an IC (CC13, IBA, Louvain-La-Neuve, Belgium) was positioned beneath the electron cone to measure bremsstrahlung radiation, serving as a surrogate for monitoring relative changes in output during UHDR delivery. All IC readings were corrected for temperature and pressure.

Before each day's animal irradiation session, 3 film measurements were acquired under the reference condition to establish the baseline DPP. Corresponding IC readings were recorded simultaneously to determine the baseline IC reading per pulse. During animal irradiation, the corresponding IC reading was compared to the baseline to determine relative output variation. Along with the baseline DPP and the output variation, we can determine actual DPP delivered to each irradiated animal, which was subsequently used in the MC engine to calculate the dose delivered for a given animal. The difference in IC readings between the solid water and rat irradiation setups was verified to be negligible.

Both the film dosimetry and IC-based output monitoring method were independently validated. EBT-XD films were selected for UHDR measurement due to their demonstrated superiority in dose rate independence[11,12]. Films were scanned 24 hours post-irradiation using a photo flatbed scanner (Epson Expression 12000XL, Suwa, Nagano, Japan), and analyzed using in-house software with a dual-channel method (green and blue)[13]. Film-measured doses were validated against the known doses delivered by the 18 MeV CONV beam in solid water, showing < 3% deviations over the 0–40 Gy dose range. The IC method was validated across 3 independent measurement sessions over a 2-week period.

## 2.4 In vivo dose and dose rate determination
### 2.4.1 MC engine

A MC engine was developed to compute in vivo dose and dose rate distributions[8]. Using the Geant4-based GAMOS MC package[14], the LINAC head geometry was explicitly simulated, and beam models were commissioned for both UHDR and CONV modes. Beam parameters including mean energy, energy spread, source emittance cone angle, and spot size were optimized to achieve <2% average absolute difference between MC-simulated and measured percent depth dose (PDD) and profiles at depths of 0–4 cm at 100 cm SSD for various field sizes. Phase-space files generated at 96 cm SSD below the electron applicator, along with animal CBCT scans, were used for subsequent in vivo dose and dose rate calculations. Animal CBCT images were imported into the Eclipse (Varian, Palo Alto, CA) for segmentation and then exported to GAMOS. Based on the tissue segmentation, proper material types were assigned to the corresponding image voxels for the simulation. The absolute doses in the MC calculation were obtained from raw values by applying a scaling factor, which was determined under the aforementioned reference condition as the ratio of the measured dose to the MC raw value. The statistical uncertainty of MC simulations was maintained below 3%.

### 2.4.2 In vivo dose distribution calculations

To streamline the workflow of our spinal cord study while maintaining accuracy, we evaluated whether the averaged relative dose distribution derived from a representative cohort of non-irradiated rats of the same age and similar weight as the experimental groups could be applied to each irradiated rat. This approach avoids the need for CBCT-based MC calculations for every individual animal. The control non-irradiated rats had undergone the standard setup applied to irradiated rats and had also been used to test setup accuracy and reproducibility, as described in Section 2.2.

For each control rat, in vivo dose distributions were calculated for both UHDR and CONV scenarios and normalized to 100% at the target point. The longitudinal profile along the C1–T2 spinal cord, the lateral profile across the target point in the L–R direction, and the $PDD_{TP}$ (PDD along the CAX normalized to the spinal cord target point) were extracted. We then evaluated whether the averaged dose distribution from this control cohort could be used for subsequent irradiated rats by assessing inter-animal anatomical variability in these profiles and $PDD_{TP}$, with dosimetric variations quantified by their standard deviations.

### 2.5 Image-guided scintillator dosimetry

A HYPERSCINT RP-FLASH scintillation dosimetry system (Medscint Inc., Quebec, Canada) was used to validate in vivo MC dose/dose rate calculations. The detector probe features a cylindrical sensitive volume of 1 mm diameter and 3 mm length and operates at a 1000 Hz sampling frequency[15], ensuring sufficient spatial and inter-pulse resolution for UHDR measurements. Prior to use, the scintillator system was calibrated under UHDR irradiation, with a dose error of <3% compared to film measurements up to 40 Gy.

A rat carcass that underwent the standard irradiation setup and CBCT imaging, was sectioned at the lumbar spine to allow scintillator insertion into the C1–T2 region. The scintillator probe, enclosed in a 6 Fr catheter, was positioned at multiple locations within the C1–T2 spinal cord, controlled by a linear translation stage (10 μm resolution, XR25P, Thorlabs, Newton, NJ). The positions were verified with the portable X-ray imaging. The scintillator measured dose and dose rate at UHDR were compared with those of MC calculations for validation.

## 3. RESULTS

### 3.1 Assessment for the setup accuracy and intra-fractional motion

For setup accuracy evaluated using the 8 control rats, the identified T2 location showed a good agreement between 2D X-ray imaging and CBCT (Table 1). The positional differences are within sub-millimeter, with mean offsets of 0.3 ± 0.3 mm (S–I) and 0.2 ± 0.1 mm (L–R). Comparisons of pre- and post-irradiation X-ray images in the 12 irradiated rats showed minimal intra-fractional T2 displacements of 0.3 ± 0.6 mm and 0.2 ± 0.1 mm in S-I and L-R directions, respectively. These results confirm the accurate 2D image-guided setup and effective immobilization, providing the foundation for high-precision irradiation delivery.

|  | Direction | Mean ± Standard deviation (mm) |
|---|---|---|
| Setup accuracy from 2D X-ray relative to CBCT. | S-I | 0.3±0.3 |
|  | L-R | 0.2±0.1 |
| Intra-fractional motion between pre- and post-irradiation X-ray image | S-I | 0.3±0.6 |
|  | L-R | 0.2±0.1 |

Table 1. The setup accuracy of the 2D X-ray image for rat spinal cord irradiation was evaluated by CBCT, while intra-fractional motion was assessed using pre- and post-irradiation X-ray images.

### 3.2 IC-based UHDR output monitoring

Across the 3 independent measurement sessions over a 2-week period, IC readings exhibited strong linearity with the number of delivered pulses within each session ($R^2 = 1$, Fig. 1e). Corresponding film measurements further confirmed the linear correlation between IC readings and delivered dose (data not shown), supporting the reliability of the IC method for output monitoring during experiment.

### 3.3 MC in vivo dose calculation
### 3.3.1 Beam data and MC model

Along the UHDR beam's CAX, the PDD at 2 x 1 cm² remained > 90% and the dose rate > 383 Gy/s within the first 2 cm depth in solid water (Fig. 2a). At 2 cm depth, the dose rate profile sustained > 209 Gy/s across the 2×1 cm² field (the grey line in Fig. 2b). These beam characteristics confirm that the 18 MeV UHDR beam provides sufficient FLASH dose rates for C1–T2 spinal cord irradiation.

For the UHDR spinal cord irradiation using a 2 × 1 cm² field, the average absolute differences between MC calculation and film measurements were 1.6% for the PDD (Fig. 2a), 1.0 % for the crossline profile, and 1.4% for the inline profile (Fig. 2b). These results confirm the accuracy of the MC beam model and support its use for in vivo spinal cord dosimetry.

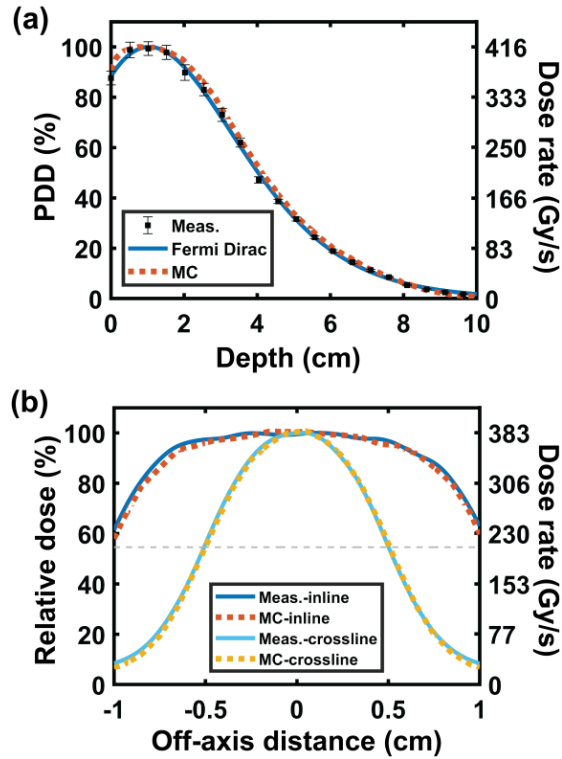

**Figure 2**. Comparison between MC-simulated and film-measured beam data for the 18 MeV UHDR beam in solid water using a 2×1 cm² field at 100 cm SSD. (a) PDD and depth dose rates along the beam CAX. The film-measured PDDs were fitted with a Fermi-Dirac distribution multiplied by a 3rd-degree polynomial with $R^2 \geq 0.99$ to generate continuous PDD curves. (b) Relative dose and dose rate profiles at 2 cm depth in both crossline and inline directions. The average absolute differences between MC simulations and measurements (Meas.) were within 2%.

### 3.3.2 Dose distribution of C1–T2 UHDR irradiation and inter-animal deviation

The MC-calculated dose distribution for a representative control rat is shown in Fig. 3a and b. Figures 3c–e present the averaged longitudinal and lateral profiles, and $PDD_{TP}$ from the 8 non-irradiated control rats, with standard deviation < 3%. This low deviation justifies the use of cohort-averaged dose distributions to represent the dose distribution of individual rats of the same age and similar weight, and confirms the reproducibility of our setup.

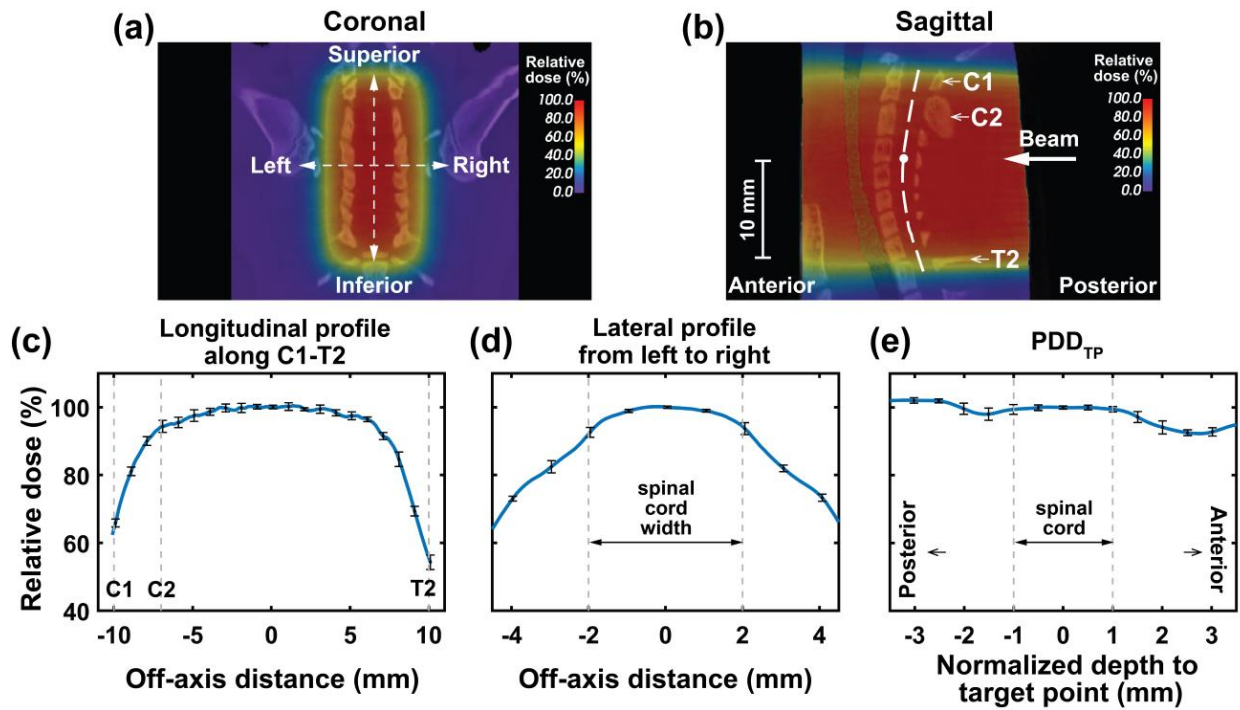

**Figure 3**. MC dosimetry of a single PA 18 MeV UHDR beam delivered to the rat C1–T2 spinal cord using a 2×1 cm² field. (a) and (b) show the dose distribution for a representative rat in coronal and sagittal views, respectively. The white dot marks the target point, located 10 mm cranial to T2 vertebra. Dose distributions are normalized to the dose at the target point. (c) Longitudinal profile along the C1–T2 spinal cord, corresponding to the white dashed curve in panel (b). (d) Lateral profile in the left-right direction. (e) $PDD_{TP}$

with the target point defined at 0 mm. Data in panels (c-e) are the average of the 8 control rats, with error bars indicating inter-animal standard deviations.

On average, the 18 MeV UHDR beam achieved uniform irradiation across a 13 mm segment of the C1–T2 spinal cord, with < 5% dose variation (Fig. 3b and c, ±6.5 mm from the 0 point in Fig. c). This avoids the dose-volume effect[7] and supports reliable assessment of the potential FLASH sparing effect. In the lateral direction, the averaged profile maintained > 91% within the ~4 mm width of the spinal cord (Fig. 3d). The averaged $PDD_{TP}$ remained > 96% within the ~2 mm across spinal cord along the posterior and anterior axis (Fig. 3e), indicating uniform dose penetration. These MC results demonstrate that shoot-through delivery using an 18 MeV beam with a 2×1 cm² field provides consistent and uniform dose coverage of the C1–T2 spinal cord in rats of the same age and similar weight.

### 3.3.3 Comparison of UHDR and CONV dose distributions

To comprehensively evaluate the effects of FLASH-RT compared to CONV-RT, MC dose distributions were also calculated for CONV irradiation and compared to those of UHDR. Figure 4a and b present the longitudinal dose profile along the C1–T2 spinal cord and the PDD along the CAX, respectively, for both beams in a representative animal. The two beams exhibited highly similar profiles and PDDs, with average absolute differences below 3%. Both UHDR and CONV beams achieved uniform irradiation across the ~13 mm C1–T2 segment (<5% non-uniformity) and maintained > 98% of the target dose throughout the spinal cord depth range (15–17 mm from rat back surface, Fig. 4b). These findings confirm that comparable dose delivery can be achieved under UHDR and CONV conditions, thereby supporting the validity of the experimental design in isolating dose rate as the primary variable for evaluating potential dose rate-dependent biological effects.

In the shoot-through irradiation geometry, the esophagus is the only organ at risk (OAR) within the beam path (Fig. 3b), receiving approximately 80% of the target dose (Fig. 4b). This suggests the potential need to monitor radiation-induced esophagitis during the follow-up period of the subsequent irradiation study.

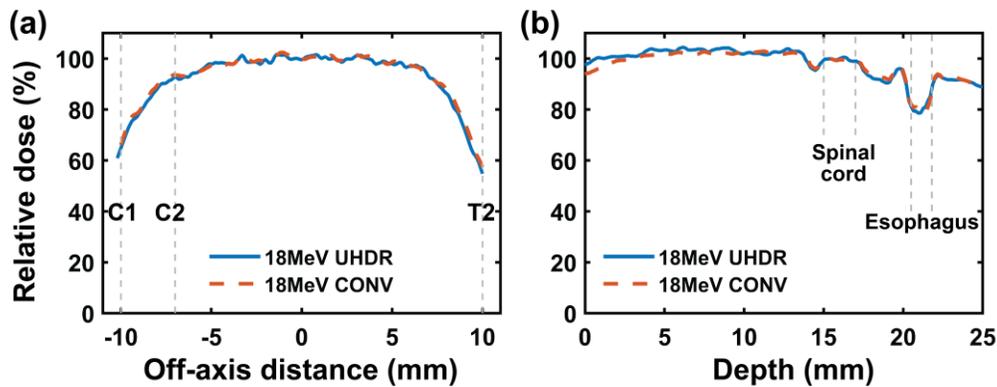

**Figure 4.** Comparison of MC dosimetry for a representative rat between the UHDR and CONV irradiation. (a) Dose profile along the C1–T2 spinal cord. (b) PDD along CAX. The posterior surface of the rat is defined at 0 mm depth.

### 3.4 In vivo MC dose and dose rate calculation validated by scintillator dosimetry

The MC-calculated dose and dose rate along the C1–T2 spinal cord were further verified with the scintillator measurements (Fig. 5a and b). The differences were < 4% within the ±10 mm off-axis distance (OAD, Fig. 5c). Considering an estimated 2.5% uncertainty in probe positioning, which arises because the linear stage measures the scintillator's curved trajectory along the spinal cord while the OAD position is projected (supplementary Fig. S1), the gamma passing rate was evaluated and reached 100% under the 2 mm/2% criteria (Fig. 5d). These results confirm the high accuracy of the MC dose engine and the precise dose delivery of the FLASH system.

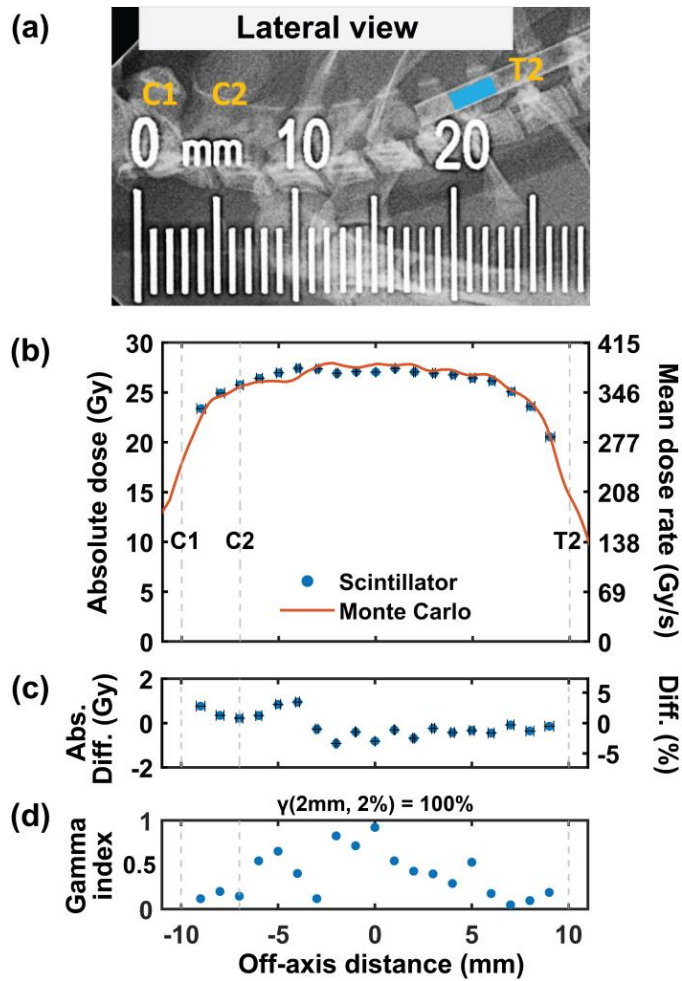

**Figure 5.** (a) X-ray imaging of the scintillator inserted into the C1–T2 spinal cord via a 6 Fr catheter. The blue square indicates the detector volume. (b) Comparison between scintillator measurements and MC calculations for the dose and dose rate along the C1–T2 spinal cord. The horizontal error bars of the scintillator measurements represent a 2.5% uncertainty in probe positioning, as the linear stage measures the scintillator's curved path along the spinal cord, while the OAD position is projected. The scintillator measurements are within standard deviation of 0.07 Gy. (c) Corresponding absolute and percentage difference in dose between measurements and MC calculations. (d) Corresponding gamma indices as a function of OAD using 2 mm/2% criteria, with a 100% passing rate.

## 4. DISCUSSION

The goals of this project are to optimize a research platform dedicated to FLASH rat spinal cord studies and to use this platform to determine whether FLASH-RT can mitigate the spinal cord toxicity compared to CONV-RT, thereby establishing its potential to mitigate late-responding tissue toxicity and assess its clinical applicability. Although the rat spinal cord has long served as a classical model for radiation-induced toxicity[7,16-20], our study introduces several key advancements going beyond traditional approaches and addressing unique challenges of UHDR in vivo irradiation. These include the image-guided spinal cord localization, real-time dose monitoring during UHDR delivery, MC dose assessment for both in vivo UHDR and CONV irradiation, and scintillator validation of MC dose and dose rate in the rat spinal cord setup.

Specifically, we first ensured accurate C1–T2 setup and localization by using the 3D-printed rat immobilization device in combination with the portable X-ray imaging (Fig. 1a-d). The 2D imaging-based localization was validated against CBCT imaging, showing submillimeter accuracy (Table 1). Intra-fractional motion was assessed by comparing T2 positions on pre- and post-irradiation X-ray images, again showing submillimeter deviation. In the S–I direction, uncertainties of this magnitude are unlikely to affect biological outcomes for the 20 mm long C1–T2 irradiation, as consistent $ED_{50}$ values have been reported across different spinal levels as long as the irradiated length exceeds 8 mm to avoid the dose-volume effect.[7,16,17]. In the L–R direction, the observed $0.2 \pm 0.1$ mm offset, arising from setup uncertainty or intra-fractional motion, corresponds to < 0.2% dose variation from the OAD = 0 position (Fig. 3d), preserving uniform spinal cord dose coverage. Together, our image-guided spinal cord setup and IC-based real-time dose monitoring (Fig. 1e) ensure accurate and reproducible C1–T2 localization and precise UHDR dose delivery.

To demonstrate the hypothesized superiority of FLASH-RT in sparing normal tissue toxicity, studies must be conducted under controlled conditions, using comparable UHDR and CONV plans and deliveries, with dose rate as the only variable. While many prior studies on spinal cord toxicity under CONV-RT have emphasized biological endpoints[16-23], few have characterized the physical dose distribution in detail[7]. Moreover, it has been recommended to report detailed dose/dose rate distributions to ensure reproducibility in FLASH research[24]. This motivated us to develop MC dose engine to evaluate 18 MeV UHDR and CONV beams. Considering this, we have developed a Geant4-GAMOS MC dose engine to provide accurate dose calculations for electron beams where we explicitly modeled the head geometry of the LINAC. For the eFLASH mode, the target and the scattering foil were removed from the LINAC geometry. The MC-calculated PDD and profiles agreed with measurements with average absolute difference < 2% (Fig.2), confirming the accuracy of the MC beam model and supporting its use for spinal cord dosimetry.

Our MC simulations further demonstrate that the 18 MeV UHDR beam achieved uniform irradiation (< 5% non-uniformity) along the central ~13 mm spinal cord segment (Fig. 3c), avoids the dose-volume effect[7]. Meanwhile it maintained > 91% and > 96% of the target dose across the spinal cord width (Fig. 3d) and along the posterior-anterior direction (Fig. 3e), respectively. These results demonstrate the uniform dose coverage of using high-energy 18 MeV electrons for rat spinal cord study. It should be noted that the spinal cord dose-response relationship for cervical-thoracic segment with CONV-RT has been extensively using photons, protons, and high-energy electrons in a shoot-through beam arrangement. The corresponding $ED_{50}$ is consistently ~20 Gy for single-fraction irradiation across all modalities (Fig. 6)[7,17-19]. This suggests that spinal cord toxicity is independent of beam modality, thereby confirming the validity of our study design of employing 18 MeV electron beams. Furthermore, our CONV beam exhibited dose distribution close to that of the UHDR beam, with difference < 3% (Fig. 4), supporting our study design in isolating the dose rate as the only variable for evaluating FLASH-dependent biological effects.

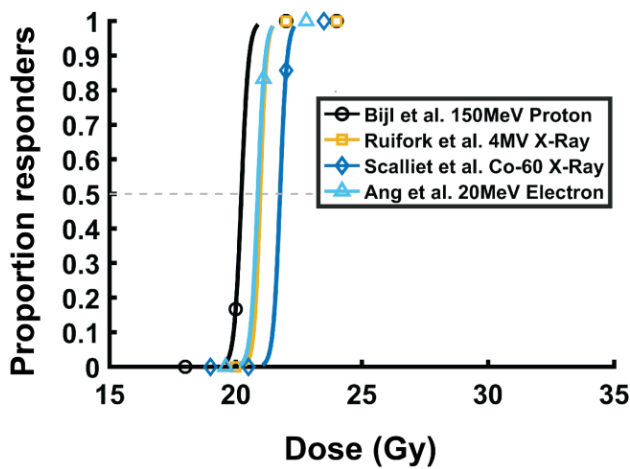

**Figure 6**. Dose-response curves of rat cervical-thoracic spinal cord subject to single fraction CONV-RT using various beam modalities, with the $ED_{50}$ consistent between 20.4 - 21.5 Gy.

When using a rat cohort of the same age and similar weight, inter-animal variability in dose profiles and $PDD_{TP}$ is minimal (<3%, Fig. 3c–e). This approach greatly simplifies the workflow and in vivo dose calculation, as the averaged dose distribution can represent that of individual irradiated rats, eliminating the need for CBCT-based MC calculations for each subject. To further validate the precision of our FLASH platform, we innovatively employed a small scintillator with a high 1000 Hz sampling frequency[15] to verify MC calculated dose and dose rate distribution in real-time during UHDR delivery within the spinal cord—representing the first such application in our field (Fig. 5). The differences between the scintillator-measured dose and MC calculations were <4%, and the gamma passing rate was 100% under the 2%/2 mm criteria. These results confirm the accuracy of the MC dose engine and demonstrate that our platform can achieve the desired dose rates and dose distributions for high-precision spinal cord studies under both UHDR and CONV conditions.

With this shoot-through delivery (Fig. 3b and 4b), the esophagus is the primary OAR and receives ~80% of the target dose. Although transient esophagitis with early weight loss (≤2 weeks) could occur, prior studies have shown rapid recovery thereafter[21-23], suggesting limited impact on study outcomes. We will monitor weight change post-irradiation of our study cohort and provide supportive care as needed. Our upcoming biological outcomes will be reported separately.

## 5. CONCLUSION

We developed the first comprehensive, image-guided preclinical platform capable of operating in both UHDR and CONV modes to investigate FLASH-RT-mitigated spinal cord toxicity using a rat model. The system integrates pulse control, real-time output monitoring, custom immobilization, precise image-guided target localization, and in vivo MC dosimetry validated against scintillator measurements. Using this platform, we can achieve accurate and reproducible dose delivery at both UHDR and CONV dose rates for rat spinal cord irradiation. This work thus establishes a robust foundation for systematic evaluation of the FLASH effect in late-responding organs and for determining clinical applicability of FLASH-RT.

## ACKNOWLEDGEMENT

We acknowledge the funding support from Cancer Prevention and Research Institute of Texas, RR200042.

## CONFLICTS OF INTEREST

The authors have no relevant conflicts of interest to disclose.